\documentclass[twocolumn,nobibnotes,showpacs,preprintnumbers,amsmath,amssymb]{revtex4}
\pdfoutput=1
\usepackage{graphicx}% Include figure files
\usepackage{dcolumn}% Align table columns on decimal point
\usepackage{bm}% bold math

\newcommand{\be}{\begin{equation}}
\newcommand{\ee}{\end{equation}}
\newcommand{\bea}{\begin{eqnarray}}
\newcommand{\eea}{\end{eqnarray}}

\newcommand{\GeV}{\; {\rm GeV}}

\newcommand{\eqnref}[1]{Eq.~(\ref{#1})}
\newcommand{\figref}[1]{Fig.~\ref{#1}}

\begin{document}
\title{Revisiting Theories with Enhanced Higgs Couplings to Weak Gauge Bosons}
\author{Spencer Chang, Christopher A. Newby, Nirmal Raj, Chaowaroj Wanotayaroj}

\affiliation{Department of Physics, University of Oregon, Eugene, OR 97403}
\begin{abstract}
Based on recent LHC Higgs analyses and in anticipation of future results we revisit theories where Higgs bosons can couple to weak gauge bosons with enhanced strength relative to the Standard Model value.  Specifically, we look at the Georgi-Machacek model and its generalizations where higher ``spin'' representations of $SU(2)_L$ break electroweak symmetry while maintaining custodial $SU(2)$.  In these theories, there is not only a Higgs-like boson but partner Higgs scalars transforming under representations of custodial $SU(2)$, leading to a rich phenomenology.  These theories serve as a consistent theoretical and experimental framework to explain enhanced couplings to gauge bosons, including fermiophobic Higgses.  We focus on the phenomenology of a neutral scalar partner to the Higgs, which is determined once the  Higgs couplings are specified.  Depending on the parameter space, this partner could have $i$) enhanced fermion and gauge boson couplings and should be searched for at high mass ($> 600$ GeV), $ii$) have suppressed couplings and could be searched for at lower masses, where the Standard Model Higgs has already been ruled out, and $iii$) have fermiophilic couplings, where it can be searched for in heavy Higgs and top resonance searches.  In the first two regions, the partner also has substantial decay rates into a pair of Higgs bosons.  We touch briefly on the more model-dependent effects of the nontrivial $SU(2)_C$ multiplets, which have exotic signals, such as a  doubly-charged Higgs.  We also discuss how the loop induced effects of these scalars tend to reduce the Higgs decay rate to photons, adding an additional uncertainty when extracting  the couplings for the Higgs boson.

\end{abstract}

\maketitle
\section{Introduction}
This year will be a key turning point in particle physics as the LHC is poised to discover the Higgs boson.  In that event, increased luminosity will enable the Higgs interactions with the Standard Model (SM) to be measured.   These crucial measurements will be the foundation for the argument that electroweak symmetry is broken by the vacuum expectation value of a fundamental scalar.   

The excitement of the latest LHC  \cite{ATLAS:2012ae, Chatrchyan:2012tx} and Tevatron \cite{FERMILAB-CONF-12-318-E} Higgs analyses is that they are all consistent with the Standard Model Higgs at 125 GeV .  There is also an interesting indication that the data prefers a Higgs boson with nonstandard values for its interactions with the Standard Model.  The statistical power of this preference is weak, but if this is confirmed with further data, this would be an enormous revolution, pointing to new physics beyond the Standard Model.  Anticipating this possibility, it is worth investigating the theoretical ramifications and correlated signals that would result for different Higgs couplings.  

One particularly interesting aspect of  recent model-independent fits to Higgs couplings is a hint that the Higgs coupling to weak gauge bosons is enhanced relative to the Standard Model \cite{Azatov:2012bz, Carmi:2012yp, Espinosa:2012ir, Ellis:2012rx, Giardino:2012ww}.  This preference is dominated by the excess in a CMS diphoton analysis channel that  isolates Higgses produced in vector boson fusion  \cite{Chatrchyan:2012tw}.   This preference is tempered by the fact that Higgs searches for $WW$ decays have not shown substantial excesses \cite{ATLAS:2012ae, Chatrchyan:2012tx}.

From a theoretical perspective, such a large   coupling to weak gauge bosons is particularly surprising since its value is crucial for unitarizing longitudinal $WW/ZZ$ scattering.  This unitarity argument suggests that the coupling cannot be larger than the Standard Model, unless there is a doubly-charged Higgs \cite{Low:2009di} which can be seen by an isospin analysis of $WW$ scattering \cite{Falkowski:2012vh}.

Thus, confirmation of enhanced Higgs coupling to electroweak gauge bosons would in itself be a sign for new physics, motivating a survey of theories that allow such enhancements.  Such theories, already considered in the past, have experimental challenges; in particular, they have potentially large precision electroweak corrections to  the $\rho$ parameter.  This is due to the need for a large contribution of electroweak symmetry breaking to come from a higher representation of $SU(2)_L$.   However, this problem can be solved by extending custodial $SU(2)$ symmetry  \cite{Susskind:1978ms, Weinberg:1979bn}, which we refer to as $SU(2)_C$, to these theories  \cite{Georgi:1985nv}.  The phenomenology of these models was studied long ago \cite{Chanowitz:1985ug, Gunion:1989ci} and a generalization of the original model is possible \cite{Galison:1983qg}, leading to an entire family of possible theories to consider.     

In light of the current data and in anticipation of future LHC Higgs results, we revisit these theories, looking for the telltale consequences given specific Higgs couplings and the  correlated signals of these models.  These theories are interesting since they serve as  theoretically and experimentally consistent phenomenological frameworks which extend the coupling parameter space explored by the model-independent fits to Higgs couplings.  

The organization of the paper is as follows:  in section \ref{sec:Theories}, we review a class of theories with enhanced Higgs couplings based on the Georgi-Machacek model; in section \ref{sec:HiggsPheno}, we explore the phenomenology of the neutral CP-even Higgs sector; in section \ref{sec:OtherSignals}, we discuss the extended Higgs scalars of these models, briefly touching upon their phenomenology and effects on the Higgs decay to photons; finally, in section \ref{sec:Conclusions}, we conclude.  We include further details about the Georgi-Machacek model in Appendix \ref{sec:GMdetails}.

\section{Theories  with Enhanced Higgs Couplings to Weak Gauge Bosons\label{sec:Theories}}
In recent years, people have outlined the key ingredients to have enhanced couplings to electroweak gauge bosons for Higgs scalars \cite{Low:2009di, Falkowski:2012vh}.  One of the necessary properties is the existence of a doubly-charged Higgs.   One such theory is well known, the Georgi-Machacek (GM) model \cite{Georgi:1985nv}.  From this example, to enhance the Higgs coupling, one needs a large amount of electroweak symmetry breaking from a higher representation of $SU(2)_L$ than the standard Higgs doublet. This naturally leads to a doubly-charged Higgs state.  However, precision electroweak constraints, in particular from the $\rho$ parameter, strongly constrain electroweak symmetry breaking from such higher representations.    The GM model avoids this by extending custodial $SU(2)$ symmetry \cite{Susskind:1978ms, Weinberg:1979bn} to this theory, naturally controlling the contributions to $\rho$.  This model was explored in depth (see for e.g. \cite{Chanowitz:1985ug, Gunion:1989ci}),  and in particular, we follow the notation in \cite{Gunion:1989ci}.

The GM model has electroweak symmetry breaking from both a standard Higgs doublet and a particular set of $SU(2)_L$ triplets  (one complex triplet with hypercharge $1$ and one real triplet with hypercharge $0$).   The custodial symmetry is manifest by writing the fields as
\bea
\phi = \left( \begin{array}{cc} \phi^{0*} & \phi^+ \\ \phi^- & \phi^0 \end{array}\right),\quad \chi = \left( \begin{array}{ccc} \chi^0 & \xi^+ & \chi^{++} \\ \chi^- & \xi^0 & \chi^+ \\ \chi^{--} & \xi^- & \chi^{0*}\end{array}\right)
\eea
where the matrices $\phi, \chi$ transform as $(2,\bar{2}), (3, \bar{3})$ under $(SU(2)_L, SU(2)_R)$.  There are four (nine) real degrees of freedom in $\phi\, (\chi)$ due to the following field relations $\phi^+ = - \phi^{-*}, \xi^+ = - \xi^{-*}, \xi^0 = \xi^{0*}, \chi^{++}=\chi^{--*}, \chi^+ =- \chi^{-*}.$  If the vacuum expectation values (vevs) of $\phi, \chi$ are diagonal, $(SU(2)_L, SU(2)_R)$ breaks down to the diagonal custodial $SU(2)_C$ symmetry.  A potential can be written down for these fields that preserves the custodial symmetry, see \eqnref{eq:potential}.  Radiative corrections can generate custodial $SU(2)$ violating terms, in particular  those due to hypercharge gauge interactions \cite{Gunion:1989ci}.  Such terms are dependent on ultraviolet physics and thus could be small in certain setups such as composite Higgs models \cite{Georgi:1985nv, Chang:2003zn}.  For the rest of this paper, we will assume such terms can be neglected, as they are required to be small due to electroweak precision constraints.  

Under this approximation, it is convenient to discuss the physical Higgs bosons in terms of custodial $SU(2)$ multiplets.  The field content under $SU(2)_C$ are two neutral singlets $H_1, H_1^{'},$ two triplets $H_3, G_3$, and a five-plet $H_5$.  $H_1^{'}$ and $H_5$ appear in $\chi$, while the $G_3$ are the eaten goldstone bosons of electroweak symmetry breaking.  This is realized by the vevs 
\bea 
\langle \phi^0 \rangle &=& v_\phi/\sqrt{2}, \quad \langle \chi^0 \rangle= \langle \xi^0 \rangle = v_\chi,\\
v_\phi &=& \cos \theta_H \, v, \; \quad v_\chi= \sin \theta_H \, v/\sqrt{8}, \nonumber
\eea
where we have defined a mixing angle for the vevs $\theta_H$. 
Gauge boson masses are generated, $m_W^2 = m_Z^2 \cos^2 \theta_W = \frac{1}{4}g^2 (v_\phi^2+8v_\chi^2)= \frac{1}{4}g^2 v^2,$ predicting $\rho =1$ at tree level as expected. For more details on the scalar spectrum, see appendix \ref{sec:GMdetails} and  \cite{Gunion:1989ci}.

In the GM model, fermion masses come from coupling to the Higgs doublet in $\phi$.  Thus, generating the SM fermion masses will put a lower bound on $\cos \theta_H$.   The couplings of the Higgs bosons to SM fields can be easily determined.  Here, we focus on the couplings for the $SU(2)_C$ singlets $H_1, H_1^{'}$.  The fermion couplings are
\bea
c_{H_1} = 1/\cos\theta_H, \quad c_{H_1'} = 0
\eea
and the couplings to $WW/ZZ$ pairs are
\bea
a_{H_1} = \cos \theta_H, \quad a_{H_1'} = \sqrt{8/3} \sin \theta_H.
\eea
Note: we have followed the convention of recent model-independent fits to Higgs couplings to normalize to the SM values, defining a fermion coupling $c = g_{h\bar{f}f}/g_{h\bar{f}f}^{SM}$ and gauge boson coupling $a = g_{hWW}/g_{hWW}^{SM}$.  Here, one sees that  the vev contributions to the $W, Z$ masses in the $\chi$ field enable $H_1'$  to have enhanced couplings to gauge bosons.  Thus, the GM model is a consistent theory where Higgs couplings to $W$ and $Z$ can exceed the Standard Model value.  Again, this is consistent with the requirement in \cite{Low:2009di, Falkowski:2012vh} since the five-plet $H_5$ has a doubly-charged Higgs.  Furthermore, due to the custodial symmetry of the model, we can have a large contribution of electroweak symmetry breaking from the vev of $\chi$.  This enables the GM model to have enhanced gauge boson couplings in an allowed region of parameter space, for $\sin \theta_H > \sqrt{3/8}$, unlike simpler theories with only a single Higgs $SU(2)_L$ triplet.   

The GM model lends itself to a simple generalization with $\phi$ and a nontrivial multiplet $\chi = (r, \bar{r})$, where $r$ is a spin $j$ representation of $SU(2)$  with $r= 2j+1 > 2$.  Such an extended breaking sector was originally noted in \cite{Galison:1983qg} and was used to generalize the GM model in \cite{Logan:1999if}.  Custodial $SU(2)$ can be extended to this generalization and the physical Higgs multiplets will be from $\phi$ ($H_1$ and $G_3$) and from $\chi$ ($SU(2)_C$ multiplets of spin $2j, 2j-1, \ldots, 0$).  This modification changes the coupling of the singlet in $\chi$ to
\bea
a_{H_1'} = \sqrt{4 j(j+1)/3} \sin \theta_H. \label{eq:ageneralization}
\eea  
Thus, larger representations used for $\chi$ lead to an even stronger coupling to gauge bosons as well as having an increasingly complicated sector of physical Higgs bosons.  

\section{Higgs phenomenology \label{sec:HiggsPheno}}
In this section, we consider the phenomenological consequences of the GM model and its generalization, focusing on the $SU(2)_C$ singlets, deferring to the next section a discussion of the nontrivial $SU(2)_C$  multiplets.  Our emphasis is on LHC signals, for the GM model's phenomenology at LEP-2 see \cite{Akeroyd:1998zr}.  In terms of the model-independent Higgs couplings $(a,c)$, the GM model is an important phenomenological framework because it extends the theoretically allowed parameter space.  In general, $H_1, H_1^{'}$ can mix, leading to  mass eigenstates
\bea
h_1 &=& \cos \alpha\, H_1 + \sin \alpha \,H_1^{'},\nonumber \\[-.4cm] \\ \nonumber
h_2 &=& -\sin \alpha \, H_1 + \cos \alpha\,  H_1^{'}.
\eea
From this mixing angle, it is easy to determine the couplings for $h_1, h_2$, which we denote by $a_{1,2}, c_{1,2}$.  Due to the current Higgs excesses and for illustration we will take $h_1$ to be the Higgs hinted at in the data, fixing its mass to 125 GeV and assuming its couplings will be measured with future data.  One can show that the physically allowed parameter space for this eigenstate is $|a_1| \leq \sqrt{8/3}$, while the GM generalization will raise the allowed range to $|a_1| \leq \sqrt{4 j(j+1)/3}$.

\begin{figure}[htb]
\begin{center}
\includegraphics[width=7cm]{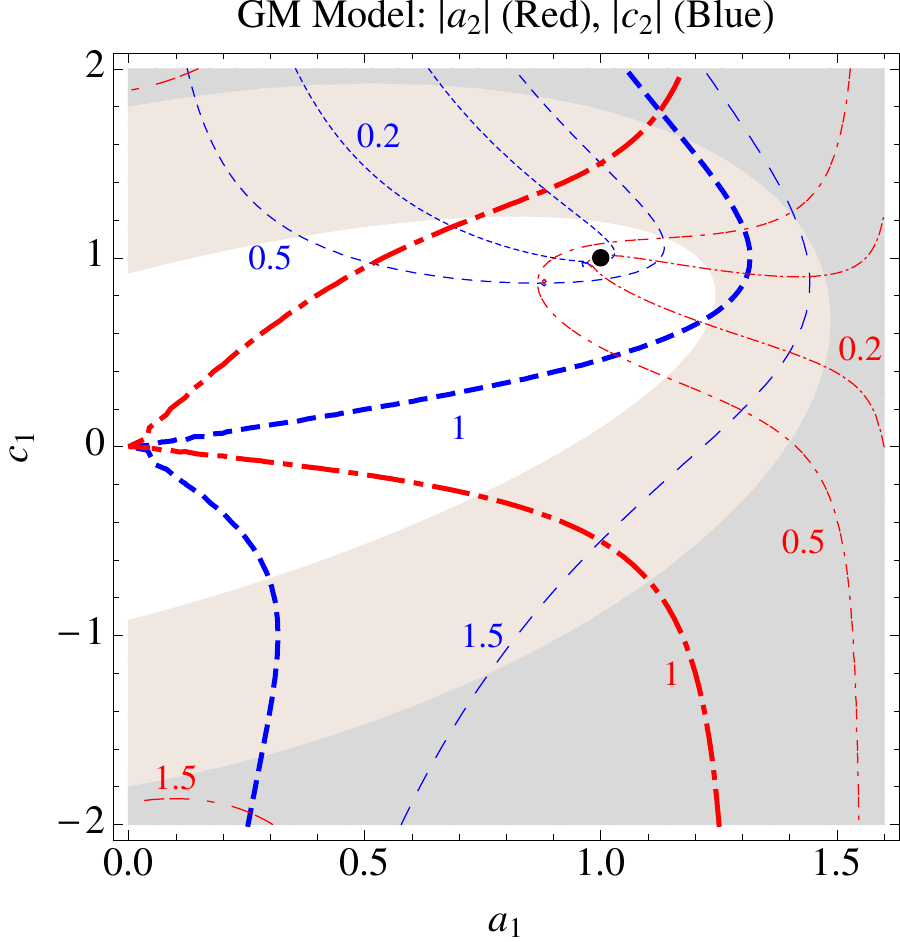}
\caption{Absolute value of couplings $a_2, c_2$ for Georgi-Machacek model as a function of the effective Higgs couplings for the $h_1$ mass eigenstate $(a_1,c_1)$.  The black dot shows the Standard Model values.  The contours are (Red, Dot-Dashed) for $a_2$ and (Blue, Dashed) for $c_2$.  The shaded contours show the excluded region from the correction to $Z\to \bar{b}b$, shown from left to right for $m_{H_3} = 200, 1000$ GeV \cite{Haber:1999zh}. \label{fig:couplings_GM}}
\end{center}
\end{figure}

Fitting to the couplings for the first mass eigenstate $(a_1,c_1)$ uniquely determines the couplings for the other eigenstate.   In \figref{fig:couplings_GM}, the absolute values of the couplings $a_2, c_2$ are shown for the GM model.  We take the absolute values for the figure presentation due to discontinuous flips of signs across the parameter space.  The relative sign of $a_2, c_2$ is important in determining $h_2$'s decay to photons and we find that there is a relative minus sign between $a_2, c_2$ only in the upper right portion of the plots (for values $c_1 > 1/a_1$), giving a constructive interference that enhances the photon decay.  On these figures, we plot constraints on $\sin^2 \theta_H$ due to modifications to the $Z\to \bar{b}b$ decay from loops involving $H_3$, which for the GM model and its generalization are  $\sin^2 \theta_H \leq 0.33\, (0.73)$ for $m_{H_3} = 200\, (1000)$ GeV \cite{Haber:1999zh}.  This constraint is plotted in \figref{fig:couplings_GM}, excluding the right side of the plots and is shown by the shaded contours in tan and gray for the two $H_3$ masses.  From the figure, one notices an interesting asymmetry between $a_2, c_2$, where $c_2$ tends to increase in magnitude as one goes to larger $a_1$, whereas $a_2$ has the opposite trend.  

The recent model-independent fits to $(a_1,c_1)$ performed by a series of papers \cite{Azatov:2012bz, Carmi:2012yp, Espinosa:2012ir, Ellis:2012rx, Giardino:2012ww} have shown that there are certain aspects of the Higgs analyses which prefer $a_1$ values larger than 1.  The discussion in \cite{Ellis:2012rx} is particularly useful for interpreting which Higgs channels are important for the data's preference in $(a_1,c_1)$.  As shown in \cite{Ellis:2012rx}, both the CMS and ATLAS photon channels prefer $a_1 > 1.$  Specifically, the CMS data prefers a region with suppressed $c_1$ ($\lesssim 0.5$) and larger $a_1$ ($\sim 1.2$), while the ATLAS data prefers an even more suppressed $c_1$ ($\sim 0$) and enhanced $a_1$ ($\sim 1.5$).  However, both the ATLAS and CMS analyses for the Higgs decay to $WW^*$ constrain this fermiophobic preference since there are no pronounced excesses.  Combining all of the channels, Ref.~\cite{Ellis:2012rx} finds that CMS has two preferred regions, one surrounding the Standard Model point with a slight preference towards enhanced $a_1$ and suppressed $c_1$ and one around $(a_1, c_1) \sim (0.8, -0.7)$; the ATLAS combination also shows a preference for the negative $c_1$ region, with a tail that extends to the fermiophobic point $(a_1,c_1) \sim (1.4, 0)$.  Thus, the LHC data is highlighting three interesting regions: $i$) near the SM values but with slight enhanced $a_1$ and suppressed $c_1$ around $(a_1,c_1) = (1.1,0.8)$, $ii$) a flipped region where $c_1$ is near $-1$ and $a_1$ slightly suppressed around $(a_1,c_1) = (0.8,-0.7)$, and $iii$) a fermiophobic region with enhanced $a_1$ around $(a_1,c_1) = (1.4, 0)$. 

A complication that will be discussed in the next section is that most of the mentioned model-independent fits to Higgs couplings assume only couplings to Standard Model particles.  In particular, the Higgs decay to photons is calculated from loop diagrams with the top quark and $W$ boson.  In the GM model and its generalizations there are additional loop diagrams due to the additional scalar content.  These must be taken into account  to determine the best fit $(a_1, c_1)$ couplings.  For now, we put aside this uncertainty, deferring details to the next section where we discuss the effects of these loops.  

\begin{figure}[htb]
\begin{center}
\includegraphics[width=7cm]{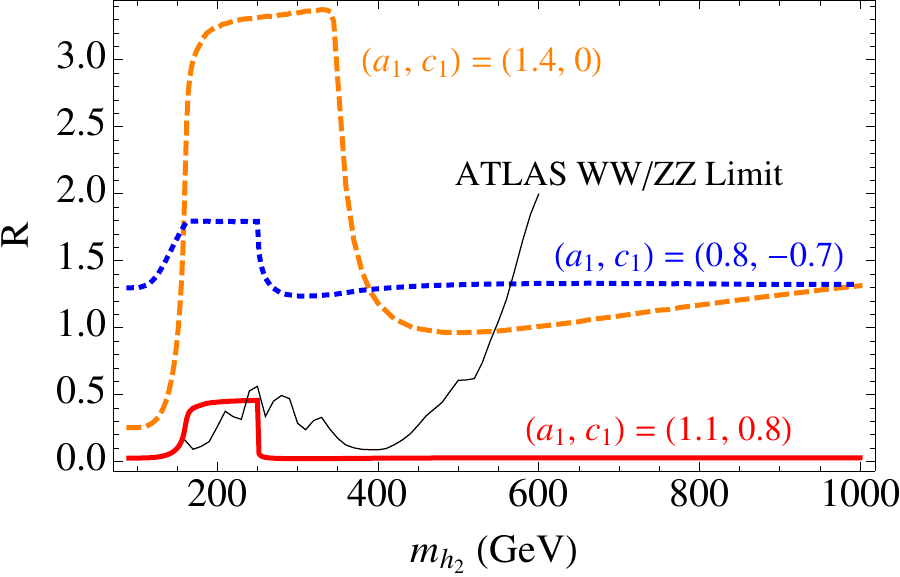}
\caption{$R$ dependence on $m_{h_2}$ for three values of $(a_1,c_1)$.  The ATLAS limit on $WW/ZZ$ Higgs searches \cite{ATLAS-CONF-2012-019} is shown in black.  \label{fig:Rmh2_GM}}
\end{center}
\end{figure}

As can be seen in \figref{fig:couplings_GM}, the GM model is able to populate a large region of the $(a_1, c_1)$ parameter space considered in these fits.  The limit from $Z\to \bar{b}b$ cuts off the large $a_1$ region, but as will be shown later, the generalizations for the GM model help to alleviate that constraint.  Notice that the GM model nicely accommodates a fermiophobic Higgs while still having perturbative Yukawa couplings to generate fermion masses (which scale as $1/\cos{\theta_H}$).  We see that the $h_2$ couplings are suppressed near the SM point and enhanced near the flipped region of negative $c_1$.  This plays a large role on the constraints and signal prospects for $h_2$.

\begin{figure*}[htb]
\begin{center}
\includegraphics[width=7cm]{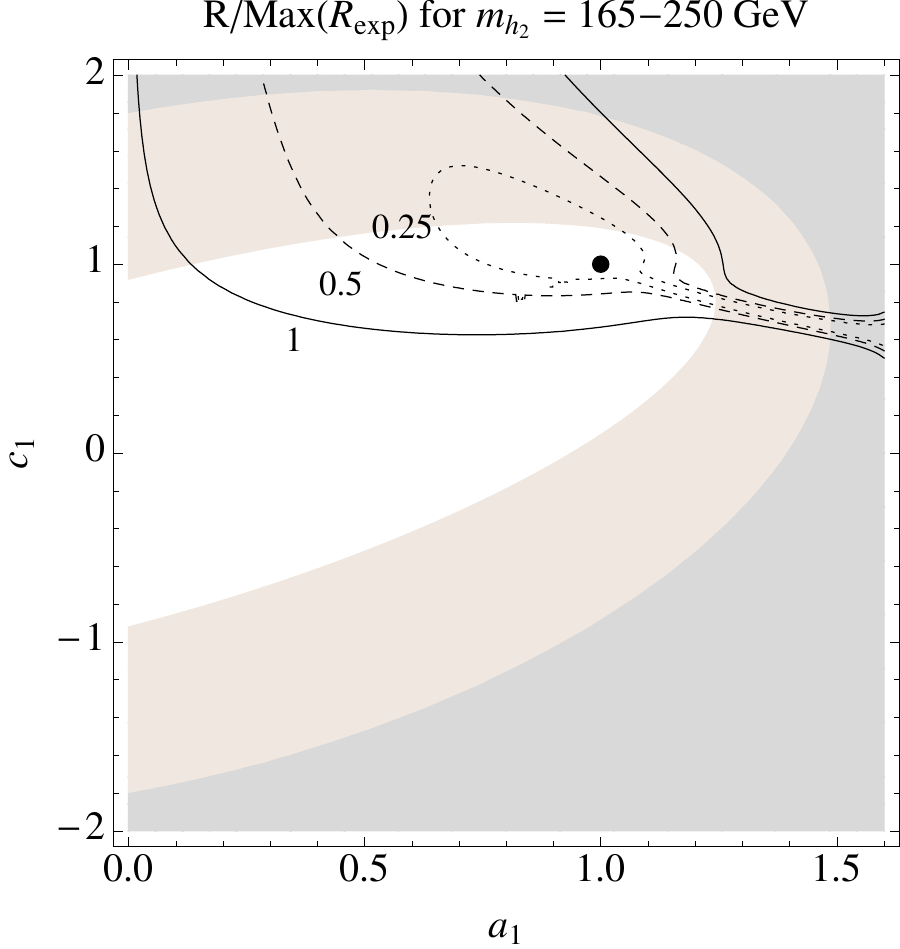}\quad \quad \includegraphics[width=7cm]{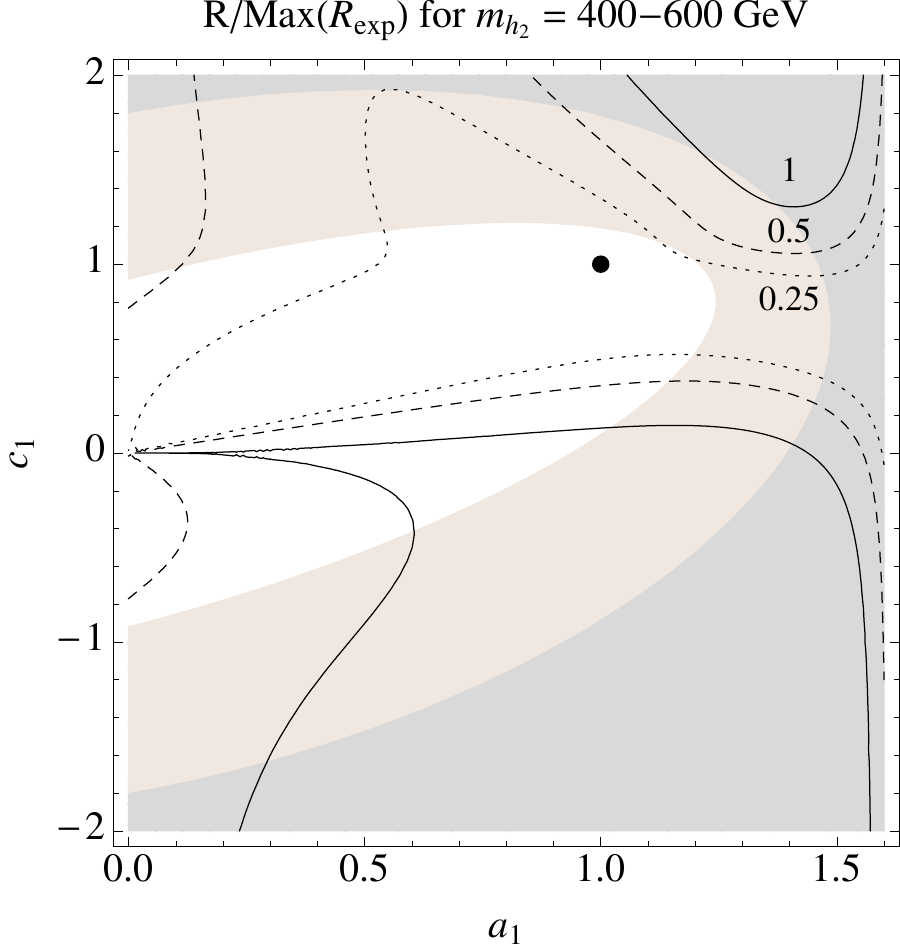}
\caption{$R/\text{Max}(R_{exp})$ for the $h_2$ mass ranges $165-250$ GeV (left) and $400-600$ GeV (right).    If the value is greater than 1, $h_2$ is not allowed in that mass range  by the LHC Higgs combined limits on $WW/ZZ$ decays.  \label{fig:Rplot_GM}}
\end{center}
\end{figure*}

We can first look at the $h_2$ phenomenology by comparing its signal rate to the Standard Model Higgs.  To simplify matters, we consider only the decays into $WW/ZZ$.  This is a useful approximation because it focuses on one number, but is  also practical since searches for a heavy Standard Model Higgs are most sensitive to these decays \cite{ATLAS:2012ae, Chatrchyan:2012tx}.  We use the ratio of rates
\bea
R = \frac{\sigma (pp \to h_2)}{\sigma (pp \to h_{SM})} \times \frac{Br(h_2 \to WW)}{Br(h_{SM} \to WW)}
\eea
where the production channel $\sigma$ is taken to be both gluon and vector boson fusion production cross sections at $\sqrt{s}= 7$ TeV LHC \cite{LHCHiggsCrossSectionWorkingGroup:2011ti}.  This assumes that the efficiencies for heavy Higgs searches are insensitive to the production mechanism and ignores the change in the Higgs width, which are good enough approximations for our purposes.    We take into account decays of the heavy Higgs into the lighter one, $h_2 \to h_1 h_1$, which is important for masses where this is kinematically open (i.e. $m_{h_2} > 2 m_{h_1} = 250 \GeV$).  This $R$ variable depends on $a_2, c_2, m_{h_2}$ and its dependence on $m_{h_2}$ is shown in \figref{fig:Rmh2_GM} for values representative of the three regions mentioned above. 

We can now compare this ratio to the current limits for the SM Higgs in the heavy mass range.  In order to see if $h_2$ is allowed in certain parts of the mass range, we look at the latest combined Higgs limits from ATLAS \cite{ATLAS-CONF-2012-019} and CMS \cite{Chatrchyan:2012tx}, using the best limit of the two as a function of mass.  Since our $R$ variable looks at $WW/ZZ$ decays, we  restrict ourselves to the mass range $165-600$ GeV where the combined limits are dominated by those decays.  The limits from the experiments fluctuate quite a bit as a function of mass, as can be seen in \figref{fig:Rmh2_GM} for the ATLAS limit \cite{ATLAS-CONF-2012-019}.  Due to the fluctuations, to get a simple understanding of what mass ranges are interesting we have to make some approximations.  First of all, $R$ is typically flat as a function of $m_{h_2}$ for a region at lower mass ($165-250$ GeV) and higher mass ($400-600$ GeV) as shown in \figref{fig:Rmh2_GM}.  This is due to the turn on of the $\bar{t}t$ and $h_1 h_1$ decays in the intermediate range.  In those two regions, we find the largest $R$ value, $\text{Max}(R_{exp})$,  allowed by both the ATLAS and CMS combinations  is  respectively $0.6$ and $1$ in the lower and higher mass region.

%\begin{figure}[htb]
%\begin{center}
%\includegraphics[width=7cm]{h2limitdirect_GM.pdf}
%\caption{The lower bound on the $h_2$ mass from the ATLAS combination \cite{ATLAS-CONF-2012-019}, using the $WW/ZZ$ decay searches in the mass range 165-600 GeV applied on the ratio $R$.  The hatched region in the lower right  is  where the constraint is outside of the considered mass range.    \label{fig:Rplot_GM}}
%\end{center}
%\end{figure}

To determine if $h_2$ is allowed in either of these two regions, we divide the average $R$ value of that range by the largest value allowed by the experiments.  This gives one an idea of how constrained an $h_2$  would be in those mass ranges.  Furthermore, one can then naively estimate how much additional luminosity would be needed to start constraining this $h_2$, since it should take a factor of $(R/\text{Max}(R_{exp}))^{-2}$ increase in luminosity from simple statistical scaling.     We plot this normalized $R$ in \figref{fig:Rplot_GM} for the two mass ranges.  Again, for values larger than 1, these plots say that that the $h_2$ cannot exist in this mass range.   As seen in the lefthand plot, only a narrow region of the $(a_1,c_1)$ parameter space allows $h_2$ in the low mass range, primarily around the SM point where the coupling $a_2$ can be suppressed.   In the righthand plot, one sees that there is a wider range allowed by the $WW/ZZ$ searches in the mass range of $400-600$ GeV for $h_2$.  The strongest  constraints are for negative $c_1$ and large $a_1$.  This reflects the fact that the $a_2, c_2$ couplings are enhanced there; in this parameter space, searches for $h_2$ with mass above 600 GeV are more motivated.  
%These two mass ranges, give a rough idea whether $h_2$ is allowed anywhere between $165-600$ GeV and if these heavy Higgs searches for $WW/ZZ$ decay searches are sensitive to it.  

\begin{figure}[htb]
\begin{center}
\includegraphics[width=7cm]{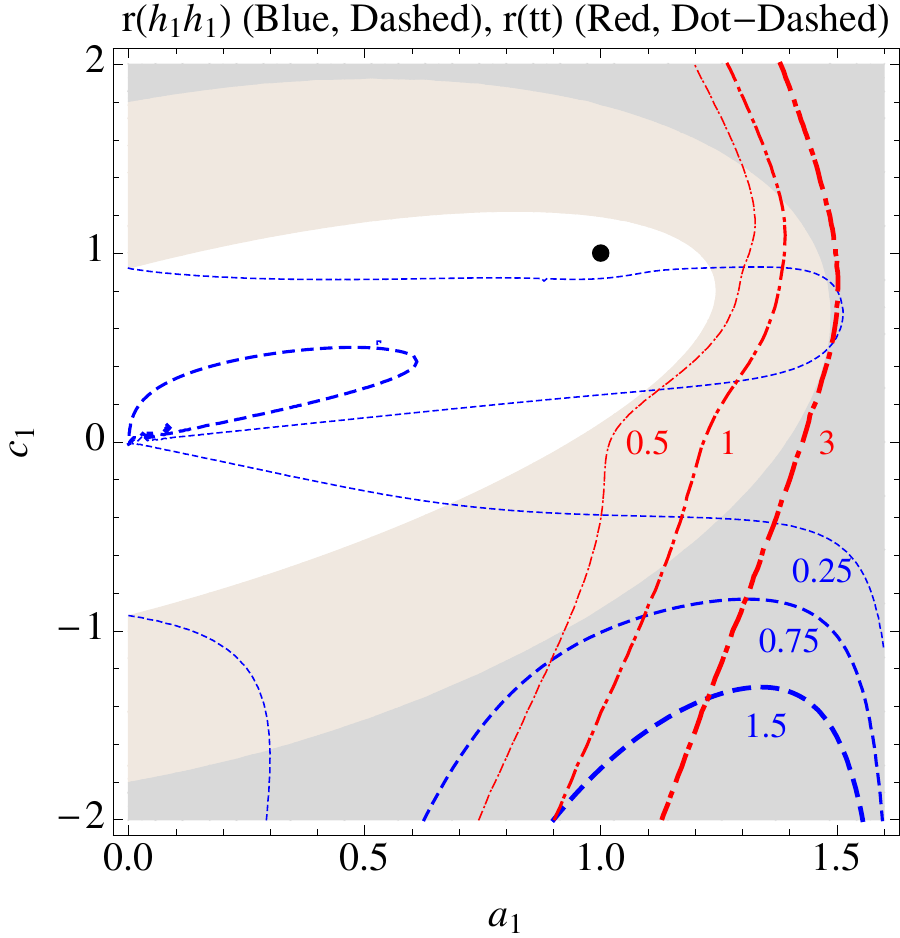} \caption{$r(X)$ of $h_2$ to $2h_1$ (Blue, Dashed) and $\bar{t}t$ (Red, Dot-Dashed) for a 400 GeV $h_2$.  This variable is large when the signal rates for these Higgs decays are large compared to the Standard Model Higgs production cross section.  \label{fig:rh2_GM}}
\end{center}
\end{figure}

In certain parts of parameter space, it could also be worthwhile to explore decays of $h_2$  into $h_1 h_1$ and $\bar{t}t$.  In \figref{fig:rh2_GM}, we plot the ratio
\bea
r(X) = \frac{\sigma (pp \to h_2)}{\sigma (pp \to h_{SM})} \times Br(h_2 \to X)
\eea
for $X= h_1 h_1, \bar{t}t$ for a $h_2$ of mass 400 GeV.  Notice that this $r(X)$ does not have a Standard Model value for the branching ratio in the denominator.  This variable $r$ is designed to determine situations where these decay signals have reasonable rates by normalizing to the SM Higgs production.  Thus, it indicates when the production of $h_2$ and the branching ratio of these modes are both large.  As can be seen in \figref{fig:rh2_GM}, for enhanced $a_1$, $h_2$ has a rate into top pairs substantially larger than  the Standard Model ($r(\bar{t}t)_{SM}=Br(h_{SM} \to \bar{t}t) \lesssim 0.2$), and thus would be interesting for top resonance searches \cite{Aad:2012wm, CMS-PAS-TOP-11-009}.   The decay into $h_1$ pairs can also have reasonable rates with $r(h_1h_1)> 0.25$, but is suppressed in the fermiophobic and $c_1 > 1$ region.  There are a variety of strategies to search for these which will depend on the branching ratios of $h_1$ but could be interesting --- for example, in $4b$ \cite{Dai:1996rn} or $2b, 2\gamma$ \cite{Baur:2003gp} signal topologies. 

\begin{figure}[htb]
\begin{center}
\includegraphics[width=7cm]{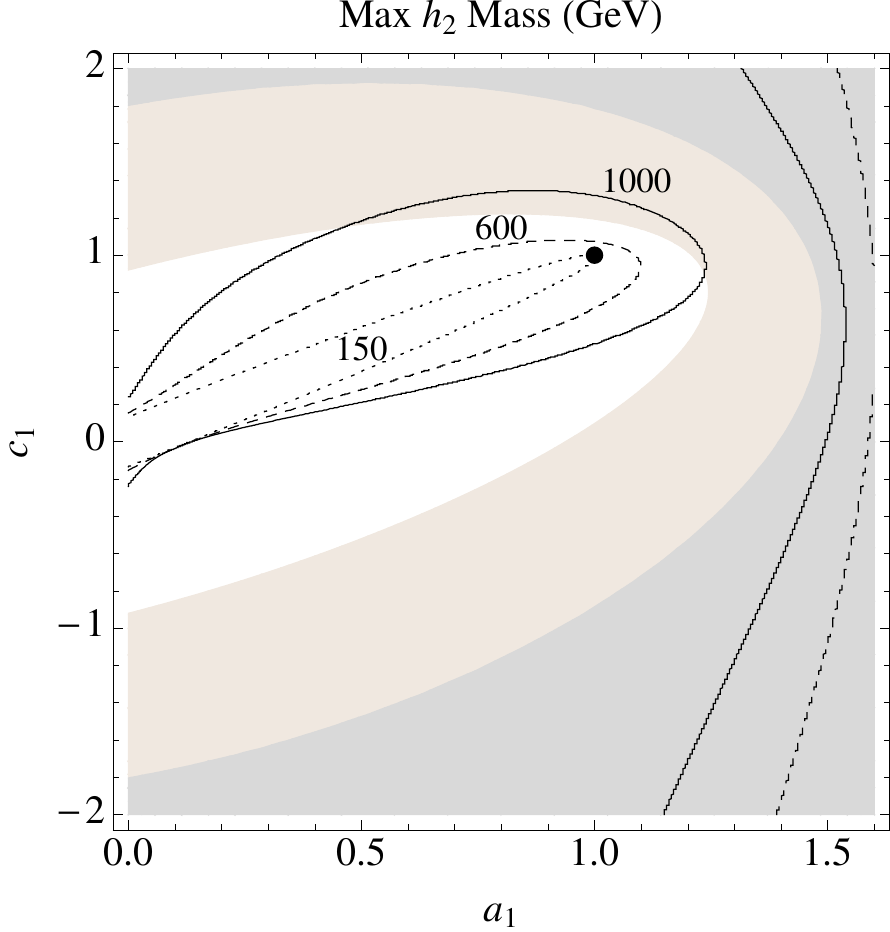}
\caption{The upper bound on the $h_2$ mass from imposing an upper limit on the quartic coupling magnitudes $|\lambda_{1,2,3}| \leq 4\pi$.  \label{fig:h2limit_GM}}
\end{center}
\end{figure}

We can also put mass bounds on $h_2$ by requiring that the quartic couplings in the potential \eqnref{eq:potential}
remain perturbative.   To illustrate this, we restrict the quartic couplings $|\lambda_{1,2,3}| \leq 4\pi$ to put upper bounds on $m_{h_2}$.  Since the masses scale as $\sqrt{\lambda} v$, for most of the parameter space this allows one to decouple $h_2$ to masses  above the existing Higgs searches ($> 600$ GeV).  However, there are some regions of $(a_1, c_1)$ whose solutions for $\theta_H, \alpha$ put more stringent upper bounds on $m_{h_2}$.  In particular, for $a_1 < 1 $, near the  $a_1 = c_1$ line, $\sin \theta_H$ approaches zero.  This puts a stringent constraint on the $h_2$ mass, since a heavy $h_2$ requires a large $\lambda_2 \sim 1/\sin^2 \theta_H$.  One can see this behavior in \figref{fig:h2limit_GM}, as the constraint is only serious around the diagonal in the upper half.  Thus, this theoretical constraint only sets a meaningful upper bound for a small fraction of the parameter space.        

\begin{figure}[htb]
\begin{center}
\includegraphics[width=7cm]{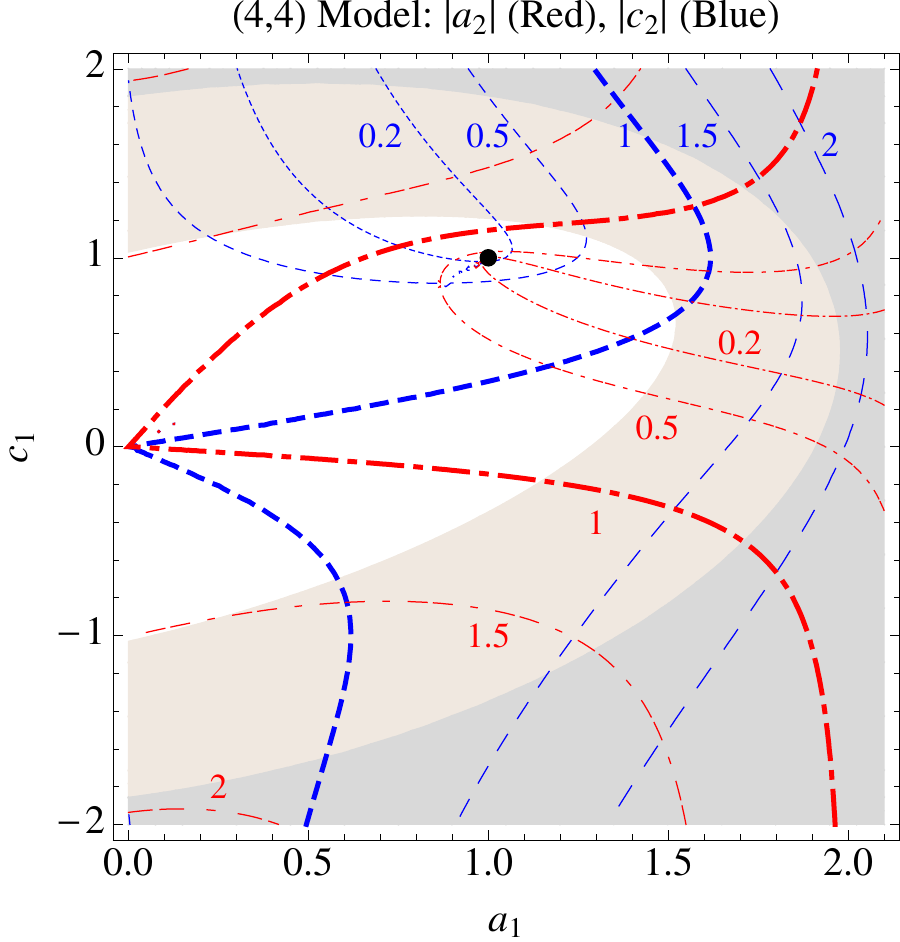}
\caption{Absolute value of couplings $a_2, c_2$ for the $(4, \bar{4})$ model as a function of the effective Higgs couplings for the $h_1$ mass eigenstate $(a_1,c_1)$.  The black dot shows the Standard Model values.  The contours are (Red, Dot-Dashed) for $a_2$ and (Blue, Dashed) for $c_2$. \label{fig:couplings_44model}}
\end{center}
\end{figure}

For generalizations of the GM model, the phenomenology of $h_2$ changes subtly, as shown in \figref{fig:couplings_44model} for the $(4,\bar{4})$ model, where the largest $a_1$ coupling allowed is increased to $\sqrt{5}$.  One sees that the $Z\to \bar{b}b$ constraint allows a larger region of $(a_1,c_1)$ coupling space.  The general behavior of the $a_2, c_2$ contours is the same, although the allowed sizes of the couplings are similarly increased.  This trend should only continue as one goes to even larger representations for $\chi$.  Since the behavior for the $h_2$ couplings are similar, the comments on the $h_2$ phenomenology apply as well to these generalizations.   We also found that the upper bound on $m_{h_2}$ from the magnitude of the quartic couplings becomes more stringent as $j$ increases, extending the region around the diagonal where it is impossible to decouple $h_2$ above 600 GeV, which improves the chances of seeing a light $h_2$.   

To summarize, the parameter space of the $h_1$ phenomenology is in one-to-one correspondence with that of  $h_2$.  One can determine general features of $h_2$'s phenomenology by combining the information from Fig.~\ref{fig:Rplot_GM} and \ref{fig:rh2_GM}.      For region $i$ of positive $(a_1,c_1)$ near the SM values, $h_2$ generally has suppressed couplings; thus in this region, LHC analyses should continue to look in mass ranges where the Standard Model Higgs has already been ruled out to dig out the suppressed rates of this partner Higgs.  For  region $ii$ of negative $c_1$ and slightly suppressed $a_1$, one sees that $h_2$ couplings to fermions and electroweak gauge bosons are generically enhanced, requiring the $h_2$ to be heavy enough to be consistent with LHC Higgs searches, $\gtrsim 600$ GeV.  However, with its enhanced rate, it would be very interesting to see updated searches for heavy Higgses that extend the  mass range.  In both these regions,  searches looking for decays into $h_1$ pairs can also be important.  Finally, in region $iii$ where $h_1$ is fermiophobic,  $h_2$ tends to be fermiophilic with enhanced couplings to fermions and with a suppressed coupling to gauge bosons.  This $h_2$ could also be observed in heavy Higgs and in   top pair resonance searches  \cite{Aad:2012wm, CMS-PAS-TOP-11-009}.

\section{More Model-dependent Signals \label{sec:OtherSignals}}
Now we will discuss  the phenomenology of the scalar fields appearing in nontrivial $SU(2)_C$ multiplets.  These multiplets can have quite exotic phenomenology due to their nonstandard quantum numbers.  For example, in the five-plet $H_5$ there is a doubly-charged Higgs.   Searches for doubly-charged Higgses at the LHC have predominantly focused on pair producing them, followed by their decays into lepton pairs \cite{ATLAS-CONF-2011-127, CMS-PAS-HIG-11-007}.  Such searches are dependent on Yukawa couplings to $\chi$ that give neutrino masses and thus are model-dependent.  However, due to the custodial symmetry in the GM model, it is also possible to get a large contribution to electroweak symmetry breaking to occur in the triplets of $\chi$, leading to a significant single production of $H_5^{++}$ via $W^+ W^+$ fusion.  Some early analyses have shown promising prospects for this to be discovered at LHC with $\sqrt{s} = 8$ TeV \cite{Chiang:2012dk} if the $\chi$ vev is large enough and  $H_5$ is light enough.  Such searches would be highly motivated if Higgs couplings to gauge bosons get a strong preference for enhancement.  The scalars in $H_3$ are very similar to the heavy Higgses of the minimal supersymmetric Standard Model, where the couplings to fermions are enhanced for large $\sin \theta_H$ and there is no coupling to gauge bosons.  However, since there is only a single Higgs doublet in $\phi$, these scalars couple universally to up and down-type fermions according to mass.  Unfortunately, the constraint from $Z\to \bar{b}b$  \cite{Haber:1999zh} tends to push the $H_3$ to masses too heavy to search for.    

The final decay products of these scalars can be even richer, since the different custodial multiplets can cascade into each other, either through $W/Z$ emission or into Higgs pairs.  In our approximation, these decays are governed by $SU(2)_C$, with decays emitting a $W/Z$  changing $j$ by 1 and the Higgs pair decays allowed if $j= j_1 + j_2$.  Incorporating $SU(2)_C$ violation would split the states within the multiplets, potentially allowing $W$ transitions if the splittings are large enough.  The generalizations of the GM model have a richer Higgs sector, given the larger content in $\chi$, leading to even more exotic charges.  However, in all of these theories, it is  possible to decouple these non-singlet custodial multiplets to masses $\sim \sqrt{4\pi} v \sim 800-1000$ GeV, which is the upper bound requiring $WW$ scattering to be unitarized perturbatively (see for e.g., \cite{Gunion:1989ci}).  A more sophisticated analysis of the GM model combining several channels gives more stringent mass limits; in particular $m_{H_1'}< 700 \GeV,\, m_{H_3} < 400 \GeV,\, m_{H_5} < 700 \GeV$ \cite{Aoki:2007ah}, constraining how much these scalars can be decoupled,  which improves the possibility of discovering these bosons.  

\begin{figure*}[t]
\begin{center}
\includegraphics[width=7cm]{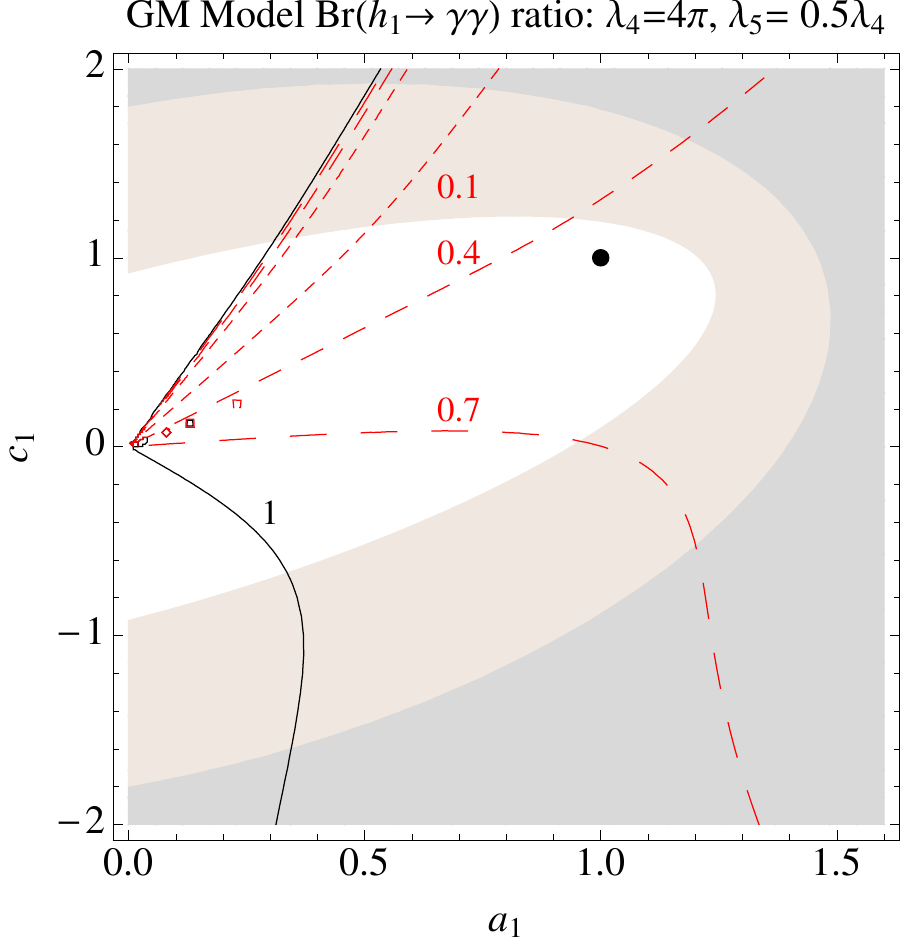} \quad  \quad \includegraphics[width=7cm]{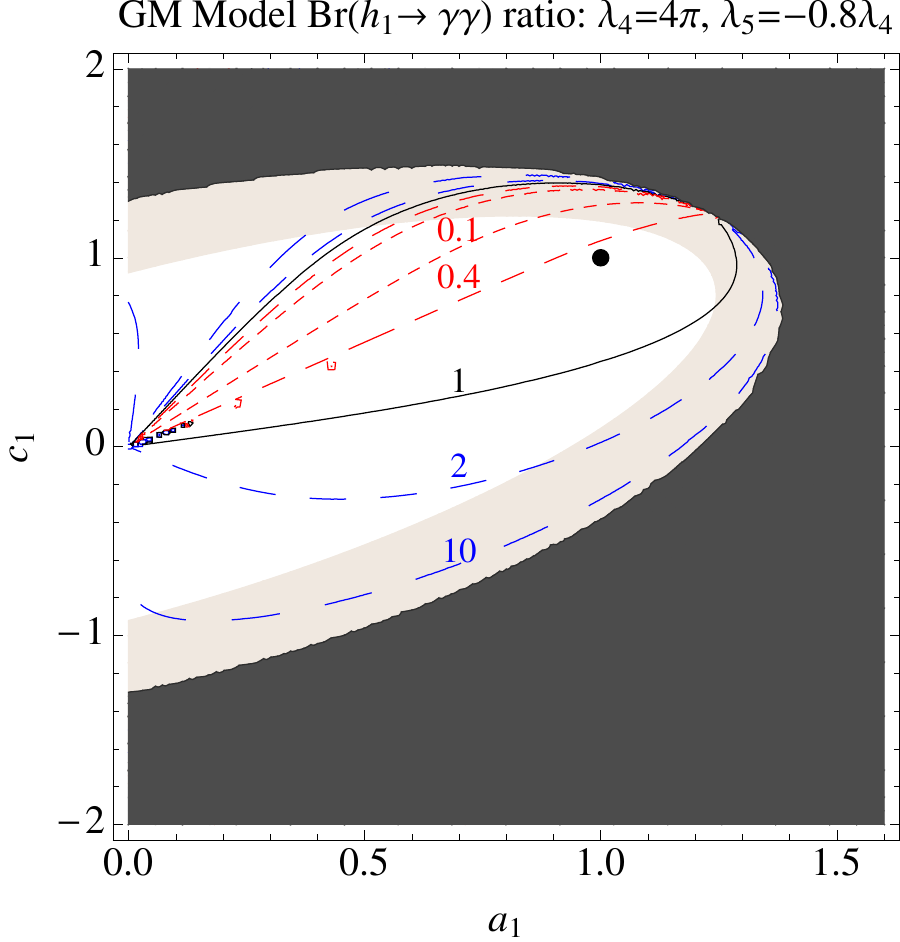}
\caption{Two example plots of the loop effects on the $h_1$ decay to photons, taking into account only the $H_3, H_5$ effects from the quartics $\lambda_4, \lambda_5$ in the potential.  The plotted contours are the branching ratio to photons with these effects accounted for divided by the branching ratio  considering just the $a_1, c_1$ couplings.  The blacked out region is where $m_{H_5} \leq 100$ GeV. \label{fig:gammaratio}}
\end{center}
\end{figure*}

Interestingly, since these additional multiplets appear in nontrivial $SU(2)_C$ multiplets, the neutral Higgses do not have to have equal couplings to $W^+ W^-$ and $ZZ$ as compared to the Standard Model.  For example,  the $H_5^0$ has $a_{WW} =\sqrt{1/3} \sin \theta_H$ and $a_{ZZ} =-\sqrt{4/3} \sin \theta_H$.  Thus, these neutral scalars can provide examples of Zphilic Higgs \cite{Farina:2012ea}, without large custodial $SU(2)$ violation.  In this scenario, the five-plet is the only viable option, since the three-plet does not couple to gauge bosons (due to its CP properties) and higher multiplets cannot couple to two gauge bosons (if $SU(2)_C$ is preserved).  Thus, in these theories, a Zphilic Higgs predicts both a doubly-charged and singly-charged Higgs with mass near 125 GeV.  For more discussion on the constraints of custodial $SU(2)$ on the allowed scalar couplings, see \cite{Low:2010jp}.

Finally, these additional Higgses have an important effect on the $SU(2)_C$ singlet phenomenology.  As discussed recently in \cite{Alves:2011kc, Akeroyd:2012ms, Wang:2012gm}, loop effects of additional charged particles will induce corrections to the $H_1, H_1'$ width into photons.  In the GM model, the charged scalars in $H_5, H_3$ tend to have contributions with the same sign as the top quark; hence, these effects tend to cancel against the $W^+$ loop, leading to a smaller decay rate into photons.  In particular, the couplings in \eqnref{eq:tripleHiggscouplings} tend to destructively interfere when the $\lambda$ couplings are positive.  To illustrate this effect cleanly, we consider the loop diagrams of the charged scalars $H_3, H_5$ that are proportional to $\lambda_4, \lambda_5$, the quartic couplings  responsible for their mass, see \eqnref{eq:GMmasses}.    The lefthand plot in \figref{fig:gammaratio} is an example of the modification to the $h_1$ diphoton branching ratio, where $\lambda_4, \lambda_5$ are both positive, demonstrating the destructive interference.  It is also possible to have negative $\lambda$'s to get constructive interference, but typically this makes the scalars lighter and risks  some of the scalars getting tachyonic masses.  This can be seen in the righthand plot in \figref{fig:gammaratio}, where $\lambda_5 = -0.8 \lambda_4$ and the black region shows where $m_{H_5} \leq 100$ GeV.  Both these plots show that to maintain the same branching ratio to photons, it is usually necessary to go to larger $a_1$ and smaller $c_1$ values.    

 In generalizations to the GM model, the higher charges of the additional Higgs states can exacerbate the interference, unless one goes to a large enough representation where the entire sign of the amplitude to photons is flipped.  However, interestingly, \eqnref{eq:generalmasses} shows that for larger representations of $\chi$, the contribution of $\lambda_5$ to the mass of the largest $SU(2)_C$ multiplet is reduced relative to $\lambda_4$.  However,  from \eqnref{eq:generalcouplings}, one sees that $\lambda_5$'s contribution to the $H_1'$ coupling to this state is not reduced.  Thus, it is easier to have negative $\lambda_5$ in the generalizations to reduce the destructive interference, while avoiding tachyonic masses for scalars.  To summarize, these loop contributions are an important effect that complicates the interpretation of the model-independent fits which for the most part include only the top and $W$ loop (however, see \cite{Ellis:2012rx}).  As an aside, we note that in a particular Higgs decay channel it is possible with enough statistics to infer the ratio of $a_1/c_1$ by looking at the rate of gluon fusion versus vector boson fusion.  Some combined fit  will eventually be able to determine the Higgs couplings to fermions, gauge bosons, {\em and} any new diagrams inducing decays to photons, which will help pin down these coupling uncertainties in these models.

\section{Conclusion \label{sec:Conclusions}}
In light of future LHC Higgs searches, we have revisited theories where the Higgs can have enhanced couplings to electroweak gauge bosons.  In particular, we reexamined the Georgi-Machacek model and its generalizations where higher ``spin'' representations of $SU(2)_L$ break electroweak symmetry while maintaining custodial $SU(2)$.  These theories widen the allowed couplings for the Higgs, serving as a consistent theoretical and experimental framework to explain enhanced Higgs couplings to $W$ and $Z$ bosons, as well as fermiophobic Higgses.  

The phenomenology of the CP-even $SU(2)_C$ singlet sector is particularly interesting, since the couplings of the two Higgses are in one-to-one correspondence.    Our assumption is that one of the scalars, $h_1$, will have its couplings to gauge bosons $(a_1)$ and fermions $(c_1)$ pinned down by future LHC analyses.  Currently, fits to LHC Higgs analyses indicate three interesting regions of $h_1$ coupling space:  $i$) near the SM values but with slight enhanced $a_1$ and suppressed $c_1$ around $(a_1,c_1) = (1.1,0.8)$, $ii$) a flipped region where $c_1$ is negative and $a_1$ is slightly suppressed around $(a_1,c_1) = (0.8,-0.7)$, and $iii$) a fermiophobic region with enhanced $a_1 \sim 1.4.$   In the Georgi-Machacek model and its generalizations, we showed that these regions have qualitatively different phenomenology for the partner Higgs boson $h_2$.   In region $i$, $h_2$ has suppressed couplings and can be searched for in lower mass Higgs searches, where the Standard Model Higgs has already been ruled out, whereas in region $ii$, it has enhanced fermion and gauge boson couplings and should be searched for at high mass ($> 600$ GeV).  In both of these regions, searches for $h_2$ decays into $h_1$ pairs are also motivated, since it can have a reasonable rate. In region $iii$, where $h_1$ is fermiophobic, $h_2$  has enhanced fermion couplings, with suppressed gauge boson couplings and thus can be picked up by both searches for heavy Higgses and top resonances.  

We also briefly discussed the model-dependent effects of the nontrivial $SU(2)_C$ multiplets, which have exotic scalar signals, such as a  doubly-charged Higgs.  Aside from direct searches, these scalars contribute nonnegligible loop effects to the Higgs decay rate to photons.  These unfortunately tend to suppress the rate and add an additional uncertainty when extracting  the couplings for the Higgs boson $h_1$.   

To conclude, if future LHC Higgs analyses indicate that the Higgs boson couplings to electroweak gauge bosons are enhanced, then it will be important to investigate theoretical frameworks that can realize such enhancements.  In this paper, we have outlined some of the important correlated signals and effects in such theories by looking at the Georgi-Machacek model and its generalizations.   There is a broad range of directions in which to test and confirm these theories, and it will take the Higgs data to determine whether nature utilizes such a  rich and complex mechanism of electroweak symmetry breaking.

\acknowledgements  
\noindent SC thanks P.~J.~Fox for useful conversations.  Part of this work was initiated at the workshop ``New Physics from Heavy Quarks in Hadron Colliders" which was  sponsored by the University of Washington  and supported by the DOE under contract  DE-FG02-96ER40956.  NR  was supported in part by the US Department of Energy
under contract number DE-FG02-96ER40969 and by the NSF under contract PHY-0918108.
\appendix 

\section{Georgi-Machacek Model Formulas \label{sec:GMdetails}}
In this appendix, we list details about the Georgi-Machacek model and its generalizations.  The potential for the Higgs fields can be written as \cite{Gunion:1989ci}
\begin{widetext} 
\bea
V&=& \lambda_1 \left(\text{Tr\,} \phi^\dag \phi -\cos^2 \theta_H v^2\right)^2 +\lambda_2 \left(\text{Tr\,} \chi^\dag \chi -\frac{3}{8}\sin^2 \theta_H v^2\right)^2+\lambda_3 \left(\text{Tr\,} \phi^\dag \phi -\cos^2 \theta_H v^2 + \text{Tr\,} \chi^\dag \chi -\frac{3}{8}\sin^2 \theta_H v^2\right)^2 \nonumber \\ \label{eq:potential} 
&& + \lambda_4 \left[\text{Tr\,} \phi^\dag \phi \, \text{Tr\,} \chi^\dag \chi -2 \sum_{ij} \text{Tr\,} (\phi^\dag \tau_i \phi \tau_j)\,  \text{Tr\,}( \chi^\dag T_i \chi T_j) \right] + \lambda_5 \left[3  \,\text{Tr\,} \chi^\dag \chi \chi^\dag \chi  -  \left(\text{Tr\,}  \chi^\dag \chi\right)^2\right]. 
\eea
\end{widetext}
Here the $\tau_i, T_i$ are the $SU(2)$ generators for a doublet and triplet.  This has a natural extension to $\chi$ of higher representation, $(r, \bar{r}) = (2j+1, \overline{2j+1})$.  This changes the factor of $\frac{3}{8}$ in $\lambda_{1-3}$  to $\frac{3}{4j(j+1)}$, the factor of $2$ in $\lambda_4$   to $\frac{4}{j(j+1)}$, and the factor of $3$ in $\lambda_5$  to $(2j+1).$  

Using this potential, for the GM model, the masses of the $SU(2)_C$ multiplets are
\bea
m^2_{H_1,H_1'} &=& \left( \begin{array}{cc} 8\cos^2 \theta_H(\lambda_1+\lambda_3) & \sqrt{6}\sin 2\theta_H \lambda_3 \\ \sqrt{6}\sin 2\theta_H \lambda_3& 3\sin^2 \theta_H(\lambda_2+\lambda_3) \end{array}\right) v^2 \nonumber \\ \label{eq:GMmasses} \\ \nonumber 
m^2_{H_3} &=& \lambda_4 v^2, \quad m^2_{H_5} = 3(\lambda_4 \cos^2 \theta_H +\lambda_5 \sin^2\theta_H) v^2.
\eea
We can also determine the Feynman rules for the triple Higgs scalar couplings.  Here, we list a few  relevant ones for $h_2\to 2 h_1$ decays and $H_1,H_1'$ decays to photons, leaving out a factor of $i$:
\bea
H_1 H_1 H_1 &=& -24 \cos \theta_H( \lambda_1 + \lambda_3 )v \nonumber \\
H_1 H_1^{'} H_1^{'} &=& -8 \cos \theta_H  \lambda_3 v \nonumber \\
H_1 H_1 H_1^{'} &=& -2\sqrt{6} \sin\theta_H  \lambda_3 v \nonumber \\
H_1^{'} H_1^{'} H_1^{'} &=& -6\sqrt{6} \sin\theta_H  (\lambda_2 +\lambda_3) v \nonumber \\
H_1 H_3^+ H_3^- &=& -8 \cos \theta_H(\sin^2 \theta_H \lambda_1 + \lambda_3 +\lambda_4)v  \label{eq:tripleHiggscouplings}\\
H_1 H_5^+ H_5^- &=& -8 \cos \theta_H\left(\lambda_3 +\frac{3}{4}\lambda_4\right)v \nonumber \\
H_1^{'} H_3^+ H_3^- &=& -2\sqrt{6} \sin \theta_H\left(\cos^2 \theta_H \lambda_2+ \lambda_3 +\frac{2}{3}\lambda_4\right)v \nonumber \\
H_1^{'} H_5^+ H_5^- &=& -2\sqrt{6} \sin \theta_H( \lambda_2+ \lambda_3 +2 \lambda_5)v. \nonumber
\eea
The couplings for the $H_5^{++}$ are the same as those of $H_5^+$ as expected from $SU(2)_C$ symmetry.  Note that we have corrected some of the  expressions in \cite{Gunion:1989ci}.

For the generalized GM model, the masses of the $SU(2)_C$ multiplets in the singlet and spin $2j$ sector are:
%\begin{widetext}
\bea
m^2_{H_1,H_1'}  = && \left( \begin{array}{cc} 8\cos^2 \theta_H(\lambda_1+\lambda_3) & \sqrt{\frac{12}{j(j+1)}}\sin 2\theta_H \lambda_3 \\  \sqrt{\frac{12}{j(j+1)}}\sin 2\theta_H \lambda_3& \frac{6}{j(j+1)}\sin^2 \theta_H(\lambda_2+\lambda_3) \end{array}\right) v^2 \nonumber  \\ \label{eq:generalmasses} \\[.1cm] \nonumber 
m^2_{H_{4j+1}} = && \frac{2(j(2j+1)\lambda_4 \cos^2 \theta_H +3 \lambda_5 \sin^2\theta_H)}{j(j+1)} v^2.
\eea
%\end{widetext}
For the triple Higgs scalar couplings, we focus on the couplings of $H_1, H_1'$ to the highest charged multiplet $H_{4j+1}$ which has a maximum charged state of charge $2j$.  Again, leaving out a factor of $i$, the Feynman rules are:
\bea
  H_1 H_{4j+1}^+ H_{4j+1}^- &=& - 8 \cos \theta_H\left[\lambda_3 +\frac{(2j+1)}{2(j+1)}\lambda_4\right]v  \nonumber \\ \label{eq:generalcouplings} \\ \nonumber 
 H_1^{'} H_{4j+1}^+ H_{4j+1}^- &=& -\frac{4\sqrt{3}}{\sqrt{j(j+1)}}\sin \theta_H( \lambda_2+ \lambda_3 +2 \lambda_5)v. \nonumber
\eea
The couplings for the other charged states in $H_{4j+1}$ are the same as those of $H_{4j+1}^+$ from $SU(2)_C$ symmetry.  
\vspace{2in}

\bibliography{EnhancedHiggs}
\bibliographystyle{apsrev}

\end{document}